\documentclass[12pt,longbibliography]{article}

\pdfoutput=1

\usepackage{cite}
\usepackage{mathrsfs,bm,fixmath,subeqn}
\usepackage{hyperref} 
\usepackage[usenames, dvipsnames]{color}
\usepackage[svgnames]{xcolor}

\evensidemargin=\oddsidemargin
\def\Eqn#1{Eq.\ (\ref{#1})}
\def\Eqs#1#2{Eqs.\ (\ref{#1}) and (\ref{#2})}
\def\3Eqs#1#2#3{Eqs.\ (\ref{#1}), (\ref{#2}) and (\ref{#3})}

\def\textcite#1{Ref.\ \cite{#1}}

\def\lag{{\mathscr L}}
\def\tilde{\widetilde}

\def\del{\partial}

\def\trev{{\mathscr T}}
\def\ccon{{\mathscr C}}
\def\ttinv#1//{\trev #1 \trev^{-1}}
\def\ccinv#1//{\ccon #1 \ccon^{-1}}
\def\ctinv#1//{\ccon\trev #1 (\ccon\trev)^{-1}}
\def\goto#1{\stackrel{#1}{\longrightarrow}}

\def\ri{{\mathrm i}}

\title{\bf Discrete symmetry transformations
  on non-abelian gauge fields}

\author{\bf Amitabha
  Lahiri$^1$\thanks{amitabha@bose.res.in}, 
  Palash B. Pal$^2$\thanks{palashbaran.pal@saha.ac.in (corresponding
    author)},
  Marina Shokova$^3$\thanks{marinamathmap@gmail.com} 
  \\
\small 1) S. N. Bose National Centre for Basic Sciences, JD Block,
Salt Lake, Calcutta 700106, India \\  
\small 2) Physics Department, University of Calcutta, 92 APC Road,
Calcutta 700009, India\\ 
\small 3) L. N. Gumilyov Eurasian National
University, Astana 010000, Kazakhstan
}

\date{}

\begin{document}

\maketitle

\begin{abstract}

   All gauge bosons of a non-abelian gauge theory do not transform the
   same way under the discrete transformations of time-reversal and
   charge-conjugation.  Moreover, the transformations rules depend on
   how the generators are chosen.  We show how well-defined rules
   pertain only to specific choices of generators, and then show how
   unified rules can be constructed, using matrix forms of the gauge
   bosons, which are completely independent of the choice of
   generators.
  
\end{abstract}

\bigskip

%%%%%%%%%%%%%%%
\section{Introduction}
%%%%%%%%%%%%%%%
The extended Lorentz group (ELG) contains the discrete symmetries of
space-inversion or parity, and time-reversal.  In addition, there is
an external automorphism of the Lorentz group in the form of charge
conjugation.  The transformation of the photon field under these
discrete symmetry operations are well-known and well-discussed
\cite{books}.  In non-abelian gauge theories, some special features
appear in the transformation of the gauge fields which are absent in
the abelian theory of quantum electrodynamics.  In this note, we
clarify the transformations of the gauge fields, as well as the
associated field strength tensors.

Needless to say, this is not a new issue.  There exist discussions on
these topics in the literature.  We want to improve on the previous
discussion in a few ways.  First, we want to show explicitly that the
transformation properties of all gauge bosons are not the same under
time-reversal and charge-conjugation.  This has been pointed out in
the context of classical solutions of Yang-Mills theories
\cite{Brandt:1979ab}, while in the context of quantum field operators
this is usually implicit \cite{Kovner:2002vt, Vafa:1984xg,
  Watson:2006yq}.  Second, we point out exactly which property of the
generators is important for writing the transformation rules,
irrespective of any particular choice of generators.  Third, we to
cast the transformation rules in a way that would be valid
irrespective of any special property, like hermiticity, of the
generators.

We set up our notation in the rest of this Introduction.  The gauge
fields appear in two different combinations in a non-abelian gauge
theory.  First, they appear through the gauge-covariant derivative,
\begin{eqnarray}
  D_\mu = \del_\mu + \ri g R^a A^a_\mu \,,
  \label{D}
\end{eqnarray}
where $R^a$'s denote the generators of the gauge group and $g$ is the
gauge coupling constant.  This combination governs the coupling of the
gauge fields to all other fields in the theory.  Second, the pure
gauge Lagrangian contains the field-strength tensor, which is the
combination
\begin{eqnarray}
  F^a_{\mu\nu} = \del_\mu A^a_\nu - \del_\nu A^a_\mu + g f^{abc}
  A^b_\mu A^c_\nu \,,
  \label{F}
\end{eqnarray}
where $f^{abc}$'s are the structure constants, defined by
\begin{eqnarray}
  [R^a, R^b] = \ri f^{abc} R^c \,.
  \label{algebra}
\end{eqnarray}
Our aim is to discuss the transformation of the gauge fields $A^a_\mu$
and the field-strength tensors $F^a_{\mu\nu}$ under the discrete
transformations of parity, time-reversal and charge-conjugation.

%%%%%%%%%%%%%%%
\section{Parity}
%%%%%%%%%%%%%%%
Parity transformation is the most straightforward among the three, so
we discuss it first.  Under parity, the spacetime derivatives
transform as follows:
\begin{eqnarray}
  \del_0 \goto P \del_0 \,, \qquad \del_i \goto P - \del_i \,.
  %\label{}
\end{eqnarray}
Therefore, the other term in the expression for the covariant
derivative must transform in the same way in order for a theory to be
parity invariant.  Since parity is a linear operator that does not
affect the generators, we conclude that
\begin{eqnarray}
  A^a_0 (t, \vec x) \goto P A^a_0 (t, -\vec x) \,, \qquad  A^a_i (t,
  \vec x) \goto P - A^a_i (t, -\vec x) \,.
  \label{PAP}
\end{eqnarray}
This is similar to the transformations obtained for the photon field
in QED: no new feature appears.

%%%%%%%%%%%%%%%
\section{Time-reversal}
%%%%%%%%%%%%%%%
New features appear for the time-reversal transformation, which we
will denote by $\trev$.  Now the derivatives change as
\begin{eqnarray}
  \del_0 \goto T \tilde\del_0 = -\del_0 \,, \qquad 
  \del_i \goto T \tilde\del_i = \del_i \,.
  %\label{}
\end{eqnarray}
To write how the covarant derivative transforms, we remember that
time-reversal is an antilinear operator, so that all coefficients
should be complex conjugated.  Thus,
\begin{eqnarray}
  \ttinv D_\mu// = \tilde \del_\mu - \ri g (R^a)^* \tilde A^a_\mu \,,
  \label{TDT}
\end{eqnarray}
where the tilde notation on the gauge fields is a shorthand of the
time-reversed fields:
\begin{eqnarray}
  \tilde A_\mu (t,\vec x) = A_\mu (-t,\vec x).
  %\label{}
\end{eqnarray}
This means that time-reversal invariance would be obtained if the
gauge fields transform in the following way:
\begin{subequations}
  \label{TAT}
\begin{eqnarray}
  \ttinv A^a_0// = \cases{ +\tilde A^a_0 & if $R^a$ is real, \cr
    -\tilde A^a_0 & if $R^a$ is imaginary,}
  \label{TA0T} \\  
  \ttinv A^a_i// = \cases{ -\tilde A^a_i & if $R^a$ is real, \cr
  +\tilde A^a_i & if $R^a$ is imaginary.} 
  \label{TAiT}
\end{eqnarray}
\end{subequations}
This feature, that different gauge bosons transform differently
\cite{Wang:2021nmi, Wan:2019oyr}, is not present in an abelian theory.
Because of the different transformation properties, any arbitrary
combination of generators may not have a well-defined transformation
property under time-reversal.

Note that we have never used any other property of the generators
except their reality.  For example, \Eqn{TAT} is valid irrespective of
any specific hermiticity property of the generators.  In fact, two
different kinds of transformation rules are required for internal
consistency.  To see what we mean, consider two gauge bosons
$W^\pm_\mu$ of an SU(2) gauge group, which couple to the ladder
generators $R^\pm \equiv (R^1\pm \ri R^2)/\surd2$, in the conventional
choice in which $R^3$ is taken to be diagonal (and therefore real),
whereas $R^1$ is real and $R^2$ is purely imaginary.  Then, using
\Eqn{TAT}, we get
\begin{eqnarray}
  \ttinv W^\pm_0 // = {1 \over \surd2} \ttinv \Big( W^1_0 \pm \ri
  W^2_0 \Big) // = {1 \over \surd2} \Big( \tilde W^1_0 \pm (- \ri)
  ({-} \tilde W^2_0) \Big) = \tilde W^\pm_0 \,,
  %\label{}
\end{eqnarray}
which is the appropriate transformation rule since the associated
generators are real.

%%%%%%%%%%%%%%%
\section{Charge-conjugation}
%%%%%%%%%%%%%%%
Let us now turn to charge-conjugation.  It is a linear operator, like
parity.  However, there is one way that it differs from the previous
two.  If we apply charge-conjugation on a charged gauge field $W^+$,
we expect to obtain $W^-$.  As we know, $W^+_\mu$ and $W^-_\mu$ are
different 4-vectors, i.e., they belong to different representations of
the ELG.  That is why charge-conjugation is called an outer
automorphism.

In order to deal with it, it is not enough to look at the covariant
derivative.  We need to consider the entire interaction term with the
gauge field.  The covariant derivative tells us that we can write the
interaction term as
\begin{eqnarray}
  \lag_{\rm int} = - J^a_\mu {A^a}^\mu \,,
  \label{JA}
\end{eqnarray}
where $J^a_\mu$ is the current.  We need its transformation property
in order to discuss the transformation property of the gauge fields.

Let us first consider the current of a multiplet of scalar fields:
\begin{eqnarray}
  J_\mu^a &=& \ri \Big( \phi^\dagger R^a \del_\mu
  \phi - (\del_\mu 
  \phi^\dagger) R^a \phi \Big) \nonumber\\
  &=& \ri \Big( \phi^\dagger_\alpha (R^a)_{\alpha\beta} \del_\mu
  \phi_\beta - (\del_\mu 
  \phi^\dagger_\alpha) (R^a)_{\alpha\beta} \phi_\beta \Big) \,,
  %\label{}
\end{eqnarray}
where in the last step, we have explicitly shown the indices
$\alpha,\beta$, which refer to different components of the multiplet
$\phi$.  Under charge-conjugation, any scalar field would transform as
follows:
\begin{eqnarray}
  \ccinv \phi// = \phi^\dagger \,.
  %\label{}
\end{eqnarray}
Thus,
\begin{eqnarray}
  \ccinv J_\mu^a // &=& \ri \Big( \phi_\alpha (R^a)_{\alpha\beta}
  \del_\mu \phi^\dagger_\beta - (\del_\mu \phi_\alpha) (R^a)_{\alpha\beta} 
  \phi^\dagger_\beta \Big) \nonumber\\
  &=& \ri \Big( {-} \phi^\dagger_\beta  (R^a)_{\alpha\beta} \del_\mu
  \phi_\alpha + (\del_\mu \phi^\dagger_\beta)  (R^a)_{\alpha\beta}
  \phi_\alpha \Big)  \nonumber\\
  &=& \ri \Big( \phi^\dagger  (-R^a)^\top \del_\mu
  \phi - (\del_\mu \phi^\dagger)  (-R^a)^\top \phi \Big) \,.
  %\label{}
\end{eqnarray}
This shows that the representation matrices in the conjugate
representation are $-(R^a)^\top$.  Note that the algebra of \Eqn{algebra}
implies that
\begin{eqnarray}
  [(-R^a)^\top,(-R^b)^\top] = \ri f^{abc} (-R^c)^\top 
  \label{algRT}
\end{eqnarray}
for any choice of generators, so the matrices $-(R^a)^\top$ constitute
a representation of the same algebra.

The same conclusion can be reached if we consider the current of
fermion fields.   In this case, the current is given by 
\begin{eqnarray}
  J^a_\mu = \bar\psi  \gamma_\mu R^a \psi  =
  (\psi^\dagger)_{\alpha A}
  (\gamma_0 \gamma_\mu)_{AB} (R^a)_{\alpha\beta} \psi_{\beta B} \,. 
  %\label{}
\end{eqnarray} 
In the last step, we have shown all indices, including the Dirac
indices in the form of $A,B$.  The charge conjugation rule for fermion
fields are as follows \cite{C}:
\begin{eqnarray}
  \ccinv \psi_A // = (\gamma_0 C)_{AB} (\psi^\dagger)_B \,,
  \label{CpsiC}
\end{eqnarray}
which also implies
\begin{eqnarray}
  \ccinv \psi^\dagger_A // = \psi_B (C^{-1} \gamma_0)_{BA} \,.
  \label{CpsidagC}
\end{eqnarray}
Therefore,
\begin{eqnarray}
  \ccinv J^a_\mu // = \ccinv (\psi^\dagger)_{\alpha A} // \,
  (\gamma_0 \gamma_\mu)_{AB} (R^a)_{\alpha\beta} \ccinv \psi_{\beta B}
  //, 
  %\label{}
\end{eqnarray}
using the linearity of the transformation.  Using \Eqn{CpsiC} and
\Eqn{CpsidagC} now, we obtain 
\begin{eqnarray}
  \ccinv J^a_\mu // &=& \psi_{\alpha E} (C^{-1} \gamma_0)_{EA} 
  (\gamma_0 \gamma_\mu)_{AB} (R^a)_{\alpha\beta} (\gamma_0 C)_{BF}
  (\psi^\dagger)_{\beta F} , \nonumber\\
  &=& -  (\psi^\dagger)_{\beta F}  (\gamma_0 C)_{BF}
  (R^a)_{\alpha\beta} (\gamma_0 \gamma_\mu)_{AB}  (C^{-1}
  \gamma_0)_{EA}  \psi_{\alpha E} \,.
  %\label{}
\end{eqnarray}
In the last step, we have merely rearranged the terms.  The minus sign
comes because of anticommutation of fermion fields.  Now we can get
rid of the Dirac indices to write
\begin{eqnarray}
  \ccinv J^a_\mu // 
  &=& -  (\psi^\dagger)_{\beta}  (\gamma_0 C)^\top 
  (R^a)_{\alpha\beta} (\gamma_0 \gamma_\mu)^\top  (C^{-1}
  \gamma_0)^\top  \psi_{\alpha} \,.  
  %\label{}
\end{eqnarray}
The definition of the matrix $C$ implies
\begin{eqnarray}
  C^{-1} \gamma_\mu C = - \gamma_\mu^\top\,, \qquad C^\top = -C \,.
  \label{CgamC}
\end{eqnarray}
Using these relations, it is easily seen that
\begin{eqnarray}
  \ccinv J^a_\mu // = \bar\psi \gamma_\mu (-R^a)^\top \psi. 
  \label{CJC}
\end{eqnarray}
Once again, it shows that the relevant representation matrices are
$(-R^a)^\top$.

So the combination given in \Eqn{JA} will be invariant under charge
conjugation now demands
\begin{eqnarray}
  \ccinv A^a_\mu // = \cases{ -A^a_\mu & if $R^a$ is symmetric, \cr
    +A^a_\mu & if $R^a$ is antisymmetric.}
  \label{CAC} 
\end{eqnarray}
Note that it is only the symmetry property that matters.  Facts like
whether the generators are real, or hermitian, are of no importance.

The different transformations for the two types of gauge bosons is
crucial in understanding what turns $W^+$ into $W^-$, which are gauge
bosons of SU(2) corresponding to the ladder generators.  In the
conventional basis in which the generator associated with $W^3_\mu$ is
taken as diagonal, the generators associated with $W^\pm$ are not
purely symmetric or antisymmetric.  Therefore, we need to trace back
to the generators which have well-defined symmetry properties in order
to obtain the charge-conjugation properties of $W^\pm$.
\begin{eqnarray}
  \ccinv W^+_\mu // = {1 \over \surd2} \ccinv (W^1_\mu + \ri W^2_\mu) //
  = {1 \over \surd2} \Big( \ccinv W^1_\mu //  + \ri \ccinv W^2_\mu//
  \Big),
  %\label{}
\end{eqnarray}
using the fact that charge-conjugation is a linear operator.  In the
representation of the SU(2) generators alluded above, $W^1$ is
associated with a generator that is real and symmetric, whereas $W^2$
is associated with a generator that is imaginary and antisymmetric.
Using \Eqn{CAC} now, we obtain
\begin{subequations}
  \label{CWC}
\begin{eqnarray}
  \ccinv W^+_\mu // 
  = {1 \over \surd2} \Big( {-} W^1_\mu  + \ri W^2_\mu \Big) = - W^-_\mu \,.
  \label{CW+C}
\end{eqnarray}
Similarly, we will get
\begin{eqnarray}
  \ccinv W^-_\mu // = - W^+_\mu \,.  
  \label{CW-C}
\end{eqnarray}
\end{subequations}
Note that it is not complex-conjugation of any number that ushers this
change.  In fact, the factor of $\ri$ is unaffected since the operation
is linear.  It is the different transformation rules of the gauge
bosons, as shown in \Eqn{CAC}, that is behind the change.

%%%%%%%%%%%%%%%
\section{The matrix notation}
\label{section-matrix}
%%%%%%%%%%%%%%%
Many authors prefer to use a matrix notation for the gauge fields.  In
other words, they consider the product of the generator and the gauge
field and treat it like a matrix field.  It will be instructive to
consider the merits of such a notation.

If we look at \Eqs{D}{TDT}, we see that the effect of time-reversal
can be summarized by saying that
\begin{eqnarray}
  \ri R^a A^a_\mu \goto T (\ri R^a)^* \tilde A^a_\mu \,.
  %\label{}
\end{eqnarray}
Thus, if we write the covariant derivative in the form
\begin{eqnarray}
  D_\mu = \del_\mu + g \bm A_\mu
  \label{DA}
\end{eqnarray}
by defining the matrix
\begin{eqnarray}
  \bm A_\mu \equiv  \ri R^a A^a_\mu \,, 
  \label{matA}
\end{eqnarray}
we can write the time-reversal transformation rule as
\begin{eqnarray}
  \ttinv \bm A_0 // = - \tilde {\bm A}_0^* \,, \qquad
  \ttinv \bm A_i // = \tilde {\bm A}_i^* \,,
  \label{TAmatT}
\end{eqnarray}
where
\begin{eqnarray}
  \tilde {\bm A}_\mu^* =  (\ri R^a)^* A^a_\mu \,.
  \label{matA*}
\end{eqnarray}
Similarly, the effect of the charge-conjugation transformation can be
summarized as
\begin{eqnarray}
  \ccinv \bm A_\mu //  = -  \bm A_\mu^\top
  \equiv - (iR^a)^\top A^a_\mu \,. 
  \label{CAmatC}
\end{eqnarray}

While these are the general rules that apply irrespective of the
choice of generators, sometimes people write these rules in a way that
apply to some specific choice that they have in mind.  For example, if
one chooses only hermitian generators, one can write
\begin{eqnarray}
  (R^a)^\top = (R^a)^* \,,
  %\label{}
\end{eqnarray}
so that the charge-conjugation rule of \Eqn{CAmatC} can be written
as~\cite{Vafa:1984xg} 
\begin{eqnarray}
  \ccinv  \bm A_\mu // = \bm A_\mu^* \,.
  \label{Cherm}
\end{eqnarray}
One can choose the hermitian generators to be either real symmetric or
imaginary antisymmetric.  We can use these properties to write
\begin{eqnarray}
  (R^a)^* = \cases {R^a & if $R^a$ is real symmetric, \cr
  -R^a & if $R^a$ is imaginary antisymmetric.}
  %\label{}
\end{eqnarray}
Combining this with \Eqn{TAT}, we can write the time-reversal rule as
\begin{eqnarray}
  \ttinv R^a A^a_0 // = R^a A^a_0 \,, \qquad
  \ttinv R^a A^a_i // = - R^a A^a_i \,.
  %\label{Therm}
\end{eqnarray}
In terms of the matrices defined in \Eqn{matA}, these equations can be
written as
\begin{eqnarray}
  \ttinv \bm A_0 //  = - \tilde {\bm A}_0 \,,
  \qquad 
  \ttinv \bm A_i //  =  \tilde {\bm A}_i \,, 
  \label{Therm}
\end{eqnarray}
accounting for an extra negative sign coming from the fact that $\ri
\goto T -\ri$.  However, it has to be remembered that \Eqs{Cherm}{Therm}
are not valid for arbitrary choice of the generators.  They use
specific hermicity properties of the generators.

Previous authors have looked at discrete transformations of gauge
field components in different contexts.  Brandt \cite{Brandt:1979ab}
considered classical solutions and found the time-reversal
transformations of gauge field components by treating them as gauge
transformations.  It appears that the rules are derived assuming a
particular type of generators.  Vafa and Witten \cite{Vafa:1984xg}
considered the quantum operators, and used the full matrix-valued
gauge field $\bm A_\mu \equiv \ri R^a A^a_\mu \equiv S^a A^a_\mu$.
Their choice of the $S^a$'s is equivalent to Hermitian $R^a$'s, and
they write the transformation rules under time-reversal and
charge-conjugation in the form of \Eqn{TAmatT} and \Eqn{Cherm} above.
While these rules are consistent with their specific choice of
generators, it should be emphasized that \Eqn{Cherm} is not valid for
a general choice: instead, one should use \Eqn{CAmatC}.  The same can
be said about the joint transformation $\ccon\trev$ \cite{Vafa:1984xg,
  Wang:2021nmi, Wan:2019oyr}.  On the other hand, Wang
\cite{PhysRevD.52.7315} wrote the time-reversal transformation rules
in the same way as Vafa and Witten, but wrote the charge-conjugation
transformation rule as $\ccinv \bm A_0 // = - \bm A_0$, $\ccinv \bm
A_i // = \bm A_i$, which does not seem to correspond to the formulas
we derived above.

%%%%%%%%%%%%%%%
\section{The field-strength tensor}
%%%%%%%%%%%%%%%
We will now see how the components of the field-strength tensor
transform under the discrete operations.  The definition of the tensor
was given in \Eqn{F}.  For parity transformation, \Eqn{PAP} gives the
rules
\begin{eqnarray}
  F^a_{0i} (t, \vec x) \goto P - F^a_{0i} (t, - \vec x) \,, \qquad
  \qquad 
 F^a_{ij} (t, \vec x) \goto P  F^a_{ij} (t, - \vec x) \,, 
  %\label{}
\end{eqnarray}
for the components of the field-strength tensor.  As for the gauge
fields themselves, these transformation rules are same as that for an
abelian theory.

The situation changes for time-reversal and charge-conjugation.  From
\Eqn{F}, we see that
\begin{subequations}
\begin{eqnarray}
  \ttinv F^a_{0i} // &=& -\del_0 \ttinv A^a_i // -
  \del_i \ttinv A^a_0 // + g (f^{abc})^* 
  \ttinv A^b_0 // \ttinv  A^c_i // \,,  \\
  \ttinv F^a_{ij} // &=& \del_i \ttinv A^a_j // -
  \del_j \ttinv A^a_i // + g (f^{abc})^* 
  \ttinv A^b_i // \ttinv  A^c_j // \,.
  %\label{}
\end{eqnarray}
\end{subequations}
Suppose now each generator is either real or imaginary.  That means
that we have a relation of the form
\begin{eqnarray}
  (R^a)^* = \xi_{(a)} R^a \qquad \mbox{(no sum on $a$)},
  %\label{}
\end{eqnarray}
with each $\xi_{(a)}$ either $+1$ or $-1$.  Then, taking the complex
conjugate of both sides of \Eqn{algebra} and using the fact that the
generators are linearly independent, we obtain the relation
\begin{eqnarray}
  \xi_{(a)} \xi_{(b)} \xi_{(c)} = - {(f^{abc})^* \over f^{abc}} \,.
  %\label{}
\end{eqnarray}
This means that, if an odd number of imaginary generators are involved
in a particular $f^{abc}$, that $f^{abc}$ will be real.  Otherwise,
$f^{abc}$ will be imaginary.  Upon complex-conjugation, the imaginary
$f^{abc}$'s will change sign.

Let us now show how the different terms in $F^a_{0i}$ would behave
under time-reversal for different combinations of the reality
properties of the generators.  
\begin{eqnarray}
  \begin{array}[t]{ccc|cccccc}
  R^a & R^b & R^c & \del_0 A^a_i &  & \del_i A^a_0 &  & g f^{abc}
  A^b_0 A^c_i  & F^a_{0i} \\
  &&&&&&& \mbox{(no sum)} & \\ \hline
  \mbox{real} & \mbox{real} & \mbox{real} & 
  (-+) && (+-) && \phantom{g} (-++) & (-) \\
  \mbox{real} & \mbox{real} & \mbox{imag} &
  (-+) && (+-) && \phantom{g} (++-) & (-) \\
  \mbox{real} & \mbox{imag} & \mbox{real} &
  (-+) && (+-) && \phantom{g} (+-+) & (-) \\
  \mbox{real} & \mbox{imag} & \mbox{imag} &
  (-+) && (+-) && \phantom{g} (-++) & (-) \\
  \end{array}  
  %\label{}
\end{eqnarray}
The table shows that, irrespective of the reality property of $R^b$
and $R^c$, one obtains the field-strength tensor to obtain a negative
sign under time-reversal if $R^a$ is real.  The case of imaginary
$R^a$ can be covered similarly, and summarized into the rule
\begin{eqnarray}
  \ttinv F^a_{0i} // = \cases{- \tilde F^a_{0i} & if $R^a$ is real, \cr
  + \tilde F^a_{0i} & if $R^a$ is imaginary.}
  %\label{}
\end{eqnarray}
Once again we emphasize that these properties do not depend on any
other characteristic of the generator.  By an exactly similar
analysis, we find
\begin{eqnarray}
  \ttinv F^a_{ij} // = \cases{+ \tilde F^a_{ij} & if $R^a$ is real, \cr
  - \tilde F^a_{ij} & if $R^a$ is imaginary.}
  %\label{}
\end{eqnarray}
If, in analogy with \Eqn{matA}, we define a matrix form of the
field-strength tensor:
\begin{eqnarray}
  \bm F_{\mu\nu} = iR^a F^a_{\mu\nu} \,,
  \label{matF}
\end{eqnarray}
we can combine both real and imaginary generators in one single rule:
\begin{eqnarray}
  \ttinv \bm F_{0i} // = \tilde {\bm F}_{0i} \,, \qquad 
  \ttinv \bm F_{ij} // = - \tilde {\bm F}_{ij} \,.
  \label{TFmatT}
\end{eqnarray}

We now perform a similar analysis for charge-conjugation.  If each
generator is either symmetric of antisymmetric, we can write
\begin{eqnarray}
  (R^a)^\top = \eta_{(a)} R^a \qquad  \mbox{(no sum on $a$)},
  %\label{}
\end{eqnarray}
where each $\eta_{(a)}$ can be $+1$ and $-1$.  Taking the transpose of
\Eqn{algebra} and comparing with the original, we now obtain
\begin{eqnarray}
  \eta_{(a)} \eta_{(b)} \eta_{(c)} = -1 
  %\label{}
\end{eqnarray}
for each nonzero $f^{abc}$.  Clearly, one cannot choose all generators
to be symmetric, although all antisymmetric generators is a
possibility.  In general, there can be some symmetric and some
antisymmetric generators.  Since charge {conjugation} is a linear
operator, it does not affect the structure constant, whether real or
imaginary.  Therefore, for symmetric $R^a$, we can make a chart of the
{possibilities} that arise for different terms of $F^{0i}$:
\begin{eqnarray}
  \begin{array}[t]{ccc|cccccc}
  R^a & R^b & R^c & \del_\mu A^a_\nu &  & \del_\nu A^a_\mu &  & g f^{abc}
  A^b_\mu A^c_\nu  & F^a_{\mu\nu} \\
  &&&&&&& \mbox{(no sum)} & \\ \hline
  \mbox{symm} & \mbox{symm} & \mbox{anti} & 
  (+-) && (+-) && \phantom{g} (+-+) & (-) \\
  \mbox{symm} & \mbox{anti} & \mbox{symm} &
  (+-) && (+-) && \phantom{g} (++-) & (-) \\
  \end{array}  
  %\label{}
\end{eqnarray}
Combining these results with similar ones obtained for antisymmetric
$R^a$, we can write
\begin{eqnarray}
  \ccinv F^a_{\mu\nu} // = \cases{-F^a_{\mu\nu} & if $R^a$ is
    symmetric, \cr
    +F^a_{\mu\nu} & if $R^a$ is antisymmetric.} 
  %\label{}
\end{eqnarray}
In the matrix notation, one can write
\begin{eqnarray}
  \ccinv \bm F_{\mu\nu} // = - (\bm F_{\mu\nu})^\top \,.
  \label{CFmatC}
\end{eqnarray}

Once again, it should be emphasized that the transformation
rules, as given in \Eqs{TFmatT} {CFmatC} are independent of the choice
of the generators.

%%%%%%%%%%%%%%%
\section{CPT, and other issues}
%%%%%%%%%%%%%%%
Generators of a Lie group form a vector space.  There are therefore
innumerable ways that a basis set of generators can be
defined.  As we showed, for time-reversal and charge-conjugation, the
transformation properties of the associated gauge bosons depend on the
choice of generators.  For time-reversal, gauge bosons corresponding
to real and imaginary generators transform differently.  If we choose
a generator that has both real and imaginary parts, the corresponding
gauge boson will not have a well-defined property under time-reversal.
Similarly, for charge-conjugation, the difference is between symmetric
and antisymmetric generators.  If a generator is a combination of the
two kinds, the corresponding gauge boson will not have a well-defined
property under charge-conjugation.

For unitary groups, one can choose hermitian generators.  These
generators can be taken as purely real or purely imaginary, as for
example the Pauli matrices for SU(2) or the Gell-Mann matrices for
SU(3).  However, it should be remembered that it is also possible to
choose all real generators, by using the diagonal generators (which
have to be real) and the combination of the other generators which act
as ladder operators.  

In the Introduction we said that the matrix notation hides the
differences between the transformation properties of different gauge
bosons.  But there is an advantage as well, coming from the fact that
the matrices $\bm A_\mu$ associated with all gauge bosons transform
the same way, so that the transformation rules written in terms of
these matrices are valid for arbitrary choices of generators, real or
imaginary, symmetric or antisymmetric, or any mixture of any kind.
This notation is therefore convenient for looking at the CPT
transformation properties of the gauge bosons.  Using the
transformation rules for $\trev$, $\ccon$ and $\mathscr P$ separately,
we get
\begin{eqnarray}
  \mathscr {CPT} \bm A_0 (t,\vec x) \mathscr {(CPT)}^{-1} &=& 
  - \mathscr {CP} \bm A_0^* (-t,\vec x) \mathscr {(CP)}^{-1} = 
  - \mathscr {C} \bm A_0^* (-t,-\vec x) \mathscr {C}^{-1} \nonumber\\
  &=& - (\bm A_0^*)^\top (-t,-\vec x)
  = - \bm A_0^\dagger (-t,-\vec x)\,.
  %\label{}
\end{eqnarray}
Similarly, for the spatial components, we get
\begin{eqnarray}
  \mathscr {CPT} \bm A_i (t,\vec x) \mathscr {(CPT)}^{-1} &=& 
  \mathscr {CP} \bm A_i^* (-t,\vec x) \mathscr {(CP)}^{-1} = 
  - \mathscr {C} \bm A_i^* (-t,-\vec x) \mathscr {C}^{-1} \nonumber\\
  &=& - (\bm A_i^*)^\top (-t,-\vec x)
  = - \bm A_i^\dagger (-t,-\vec x)\,.
  %\label{}
\end{eqnarray}
In summary, then, we can write
\begin{eqnarray}
  \mathscr {CPT} \bm A_\mu (t,\vec x) \mathscr {(CPT)}^{-1} &=& 
  = - \bm A_\mu^\dagger (-t,-\vec x)\,.
  %\label{}
\end{eqnarray}
This is necessary for the CPT theorem to hold.

In conclusion, we demonstrated that different gauge bosons of a
non-abelian gauge theory can transform differently under time-reversal
and charge-conjugation, depending on the way the generators are
chosen.  We also showed how the matrix-valued gauge bosons can give
the transformation rules in a way that does not depend on the choice
of the generators.  We also showed that these transformation rules,
written in some specific ways using some specific choice of
generators, are not valid in general, and indicated what the
representation-independent rules should be.

%%%%%%%%%%%%%%%
\section*{Acknowledgement}
%%%%%%%%%%%%%%%
M.S. thanks Edward Witten and Juven Wang for email correspondences.

%%%%%%%%%%%%%%%%%%%%
\bibliographystyle{h-full.bst}
\bibliography{ptcbib}

\begin{thebibliography}{1}
\bibitem{C}
Some texts use expression which are specific to a particular choice of the
  Dirac matrices. Some books use $C\gamma_0^\top$ in place of $\gamma_0C$,
  which is the same, apart from an irrelevant overall sign, because of
  \Eqn{CgamC}.
\end{thebibliography}

\end{document}